\documentclass[prb,preprint,showpacs,preprintnumbers,amssymb]{revtex4}

\usepackage{graphicx}
\usepackage{amssymb}
\usepackage{mathrsfs}
\usepackage{dcolumn}
\usepackage{bm}
\begin{document}

\title{The Phase Diagram of an Anisotropic Potts Model}
\author{M. R. Ahmed}
\email[Corresponding author: ]{php02mra@shef.ac.uk}
\author{G. A. Gehring}
\affiliation{Department of Physics and Astronomy, University of Sheffield, Hicks Building, Hounsfield Road, Sheffield, S3 7RH, UK.}

\begin{abstract}
A study is made of an anisotropic Potts model in three dimensions where the coupling depends on both the Potts state on each site but also the direction of the bond between them using both analytical and numerical methods. The phase diagram is mapped out for all values of the exchange interactions. Six distinct phases are identified. Monte Carlo simulations have been used to obtain the order parameter and the values for the energy and entropy in the ground state and also the transition temperatures. Excellent agreement is found between the simulated and analytic results. We find one region where there are two phase transitions with the lines meeting in a triple point. The orbital ordering that occurs in $LaMnO_3$ occurs as one of the ordered phases.
\end{abstract}

\pacs{05.10.Ln,75.30.Kz,05.70.Fh,64.60.Cn}

\maketitle

\section{\label{introduction}Introduction}
	There have been many studies of the phase transitions of the Potts model\cite{Potts52} (for a general review see Wu 1982 \cite{Wu82}), largely due to the richness of its physical content and its relevance in real physical systems\cite{Domany77}. While a large body of exact and rigorous results are now known, a number of problems particularly those associated with models with antiferromagnetic and multi-site interaction are still being investigated.

	The ferromagnetic, FM, case has been well studied\cite{Potts52,Straley73,Zia75,Blote79} and it is now believed that the $q$-state FM Potts model in 3d for $q\geq 3$ exhibits a first-order transition\cite{Blote79}. In systems which order ferromagnetically, it is known that the critical behaviour of the system near the critical temperature $T_C$ is not affected by the nature of the underlaying lattice. The critical behaviour depends only on the dimensionality $d$ and number of components of the order parameters.

	Antiferromagnetic, AF, Potts models have been shown to possess interesting and unusual properties. The ground state entropy is nonzero whenever the number of spin states is $q>2$. The $q=3$ model on square lattice has critical points only at zero temperature \cite{Baxter82,Nightingale82,Jayaprakash82,Temesvari82,Fucito83}. In three dimensions, the evidence indicates the existence of a phase transition for $q=3,4$ and 5 that is believed to be weakly first order, although the nature of these transitions has been uncertain\cite{Banavar80,Hoppe86,Kotecky85,Ono86,Marques88,Ueno89} and a finite entropy\cite{GGrest81} is found at $T=0$. The effects of frustration in the Potts model have been studied by considering competing nearest and next nearest neighbour interactions\cite{Caracciolo88}.

	A three state Potts model in three dimensions allows for the possibility that the Potts states and the space coordinates are coupled.  This arises physically where there is strong Jahn-Teller coupling of an electronic doublet typically from $d$-electrons coupled to the two dimensional lattice distortions with strong unharmonic terms as discussed by Kanamori \cite{Kanamori61}.  This leads to three orbits which are used as the three states of the Potts model,
\begin{eqnarray}
&|x\rangle=(3x^2-r^2)f(r),        \nonumber \\
&|y\rangle=(3y^2-r^2)f(r),         \nonumber \\
&|z\rangle=(3z^2-r^2)f(r),
\end{eqnarray}

where $f(r)$ is a radial function and $r^2=x^2+y^2+z^2$.

\begin{figure}[h!]
\includegraphics[scale=0.5]{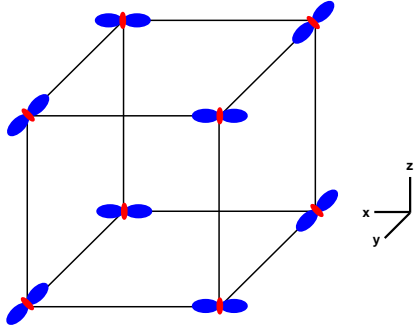}
\caption{\label{OO}The orbital ordering in $LaMnO_{3}$. It is antiferromagnetic in x-y plane and ferromagnetic along the $\hat{z}$ direction.}
\end{figure}

In such a model the interaction between orbits depends on both the type of orbit and the direction of the bond between them. The compound $LaMnO_3$ has orbital order of this type, as seen in Fig. \ref{OO}, and the interactions between the orbits have been calculated\cite{Mizokawa99,Khomskii03,Farrell04}.

\begin{figure}[h!]
\centering
\includegraphics[scale=0.8]{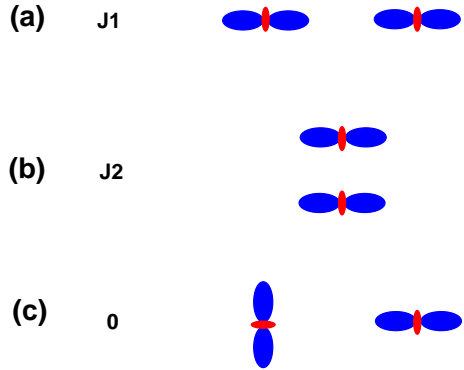}
\caption{\label{Jtype1}The types of the orbital interaction used in our simulation (a) $J_1$ refers to head-to-head ordering in one direction, (b) $J_2$ effect is to order the same states in parallel to form layers in two directions, (c) orthogonal ordering is considered as a zero interaction.}
\end{figure}

	We define the anisotropic three dimensional Potts model in terms of two interactions, $J_1$ and $J_2$ indicated in Fig. \ref{Jtype1}. The interaction $J_1$ is for two sites occupied by the same orbit, say $x$ where the bond between them is along $\hat{x}$ direction (a head to head configuration), $y$ orbits in $\hat{y}$ direction or $z$ orbits in $\hat{z}$ direction as shown in Fig. \ref{Jtype1}a. The interaction $J_2$ is when the orbits, say $x$, are separated by a bond in one of the other directions (a side to side configuration) $\hat{y}$ direction or $\hat{z}$ direction, $y$ separated by a bond in $\hat{x}$ direction or $\hat{z}$ direction or $z$ separated by a bond in $\hat{x}$ direction or $\hat{y}$ direction as shown Fig. \ref{Jtype1}b. Finally, any two orthogonal orbits have zero exchange interaction as shown Fig. \ref{Jtype1}c.
\vspace{1cm}

	In this paper we investigate the phases over the whole $J_1$-$J_2$ plane. We find an extraordinarily rich phase diagram. There are no less than six distinct phases and special critical properties on the lines separating them. In the case where $J_1=J_2=J$ we recover the results of the isotropic three dimensional ferromagnetic Potts models (for $J>0$) and the antiferromagnetic Potts models (for $J<0$). Also, the ordering of $LaMnO_3$ occurs as one of the phases.  We find that all of the phases except the ferromagnetic and isotropic antiferromagnetic phase are frustrated as the Potts states on all the sites cannot be arranged to optimize all the interactions.

	We use Monte Carlo simulations to identify the ordering that occurs in the low temperature limit in each of the phases and find the energy and entropy of the ground state of each phase.  The simulated values for the specific heat are used to find the variation of the transition temperature over the $J_1$-$J_2$ plane and to investigate qualitatively the order of the transition by comparing the form of the specific heat anomalies observed with the well-studied cases of the ferromagnetic and antiferromagnetic Potts model. The methodology is described in section 2.  In section 3 the results are presented for the six phases. Results for the phase boundaries are given in section 4. Finally the conclusions are given in section 5.

\section{Methodology}
\subsection{The Model}
	The Hamiltonian for the standard (isotropic) Potts model Eq. (\ref{iso}) and the anisotropic three state Potts models Eq. (\ref{aniso}) on a simple cubic lattice are given below.  The factor of 1/2 is included to correct for double counting.
\begin{equation}
H_{IS}=-\frac{J}{2}\sum_{\langle i,j\rangle}^{N}\delta_{S_{i},S_{j}},
\label{iso}
\end{equation}
where $S_{i}=x,y$ or $z$ is one of the three states on site $i$, ${\delta}_{S_{i},S_{j}}$ is the Kronecker delta which equals 1 when the states on sites $i$ and $j$ are identical, ${S_{i}}$ = ${S_{j}}$, and is zero otherwise, $\langle i,j\rangle$ means that the sum is over the nearest neighbour pairs, $J$ is the integral exchange and $N$ is the total number of the sites in the lattice. The anisotropic Potts model, that is studied here, differs from the standard Potts model with Hamiltonian given by Eq. (\ref{iso}) because the exchange interaction depends both on the orbit and on the direction of the bond $\underline{\rho}$,
\begin{equation}
H_{AIS}=-\frac{1}{2}\sum_{\langle i,j\rangle}^{N}J_{S_i} (\underline {\rho}_{ij}) \delta_ {S_{i}, S_{j}},
\label{aniso}
\end{equation}
where $\underline{\rho}_{ij}=\underline{R}_i-\underline{R}_j$.

	There are two exchange interactions for the anisotropic model. The 'head to head' interaction $J_1$ is defined by $J_{x}(\underline{\rho}) = J_1$ for $\underline{\rho}=\pm \hat{x}a$, $J_{y}(\underline{\rho}) = J_1$ for $\underline{\rho}=\pm \hat{y}a$ and $J_{z}(\underline{\rho}) = J_1$ for $\underline{\rho}=\pm \hat{z}a$. The 'side to side' interaction $J_2$ is defined by $J_{x}(\underline{\rho})=J_2$ for $\underline{\rho}=\pm \hat{y}a$ or $\underline{\rho}=\pm \hat{z}a$, $J_{y}(\underline{\rho})=J_2$ for $\underline{\rho}=\pm \hat{x}a$ or $\underline{\rho}=\pm \hat{z}a$ and $J_{z}(\underline{\rho})=J_2$ for $\underline{\rho}=\pm \hat{x}a$ or $\underline{\rho}=\pm \hat{y}a$. Thus each site has a coupling $J_2$ to four neighbours and a coupling $J_1$ to two neighbours. This is shown in Fig. \ref{Jtype1}. It is worth mentioning that these types of interaction do not affect the overall cubic symmetry of the lattice.

For each phase we follow the procedures below.
\begin{enumerate}
\item We perform a Monte Carlo simulation to find the nature of the ground state order. A phase is defined as a region that exhibits the same configuration in the ground state. We identify the nature of the long range order and the fraction of the sites that form the ordered state.
\item From the observed ground state configuration we obtain an analytic expression for the ground state energy per site $u_n(0)$ and for the ground state entropy per site $s_n(0)$ if possible, where $n$ is the phase number. The energy, $\Delta u_n$, which is the ordering energy of the phase is obtained as $\Delta u_n=u_n(\infty)-u(0)$ for each phase.
\item We compare the analytic values of the ground state energy and entropy deduced from the observed order with those from the Monte Carlo simulations for each phase.
\item We confirm that the phase line between two regional phases with different ground state order occurs for values of $J_1$ and $J_2$ such that the ground state energies of the two phases are equal.
\item The Monte Carlo simulations are used to estimate the transition temperatures, $T_C$, around the phase digram and the nature of phase transition (first or second order) is determined in some cases.
\item We use a combination of the Monte Carlo results and analytic results to obtain information about the type of order and ground state entropy, $s_{L_n}(0)$, occurring on the boundary lines.
\end{enumerate}

\subsection{Monte Carlo Simulations}
Our Monte Carlo calculations have been carried out on 3d finite cubic lattices (with linear size $L=8$) with periodic boundary conditions. All our simulations have made use of the Metropolis algorithm with the spin being chosen at random, and with averaging performed over $10^5$ Monte Carlo steps per site. In most of the phase diagram this gave clear results. Where convergence was slow we checked that we had found the true ground state by both increasing the number of Monte Carlo steps and also by looking at a $L=10$ lattice as explained below. Results at low temperatures were obtained by cooling down from a high-temperature random configuration as discussed by Banavar et al\cite{Banavar82}.

	The internal energy per site of the system obtained from the simulation is given as follows,
\begin{equation}
u(T) =\frac{1}{N}\langle H_{AP}\rangle_T,
\label{eneavg}
\end{equation}
where $N=L^3$. We checked that $u(\infty)=-\frac{1}{3}(J_1+J_2)$ as expected from a random array of Potts model.

	The specific heat $c_V$ per site can be obtained from the energy $u$ as follows,
\begin{equation}
c_V=\frac{1}{Nk_BT^2}(\langle u^2\rangle-\langle u\rangle^2),
\label{eqCv}
\end{equation}
where $\langle u^2\rangle$ and $\langle u\rangle ^2$ are the average and squared average over MC steps for $u^2$ and $u$ respectively. The entropy per site, $s(T)$, is obtained from integration over the specific heat where $s(T_1)$ and $s(T_2)$ are the entropy at the lower and higher temperatures respectively. We chose $T_2$ in the high temperature limit where we can take $s(T_2)=k_Blog_e3$ and used the simulation to evaluate the entropy in the ground state, $s(0)$.

\begin{equation}
 s(T_2)-s(T_1)=\int^{T_2}_{T_1} \frac{c_V}{T}dT.
\label{eqS}
\end{equation}
The errors in the determination of the ground state entropy are estimated as follow: (i) by comparing the simulated values with the exact result obtained from the order parameter, (ii) for the antiferromagnetic phase we compare with published data, (iii) finding the change in $\Delta S$ that result is from changing the parameters in the simulation such as the temperature steps, $\Delta T$, the MC steps or the lattice linear size, $L$.
\vspace{0.1cm}
\begin{figure}[h!]
\centering
\includegraphics[scale=0.3]{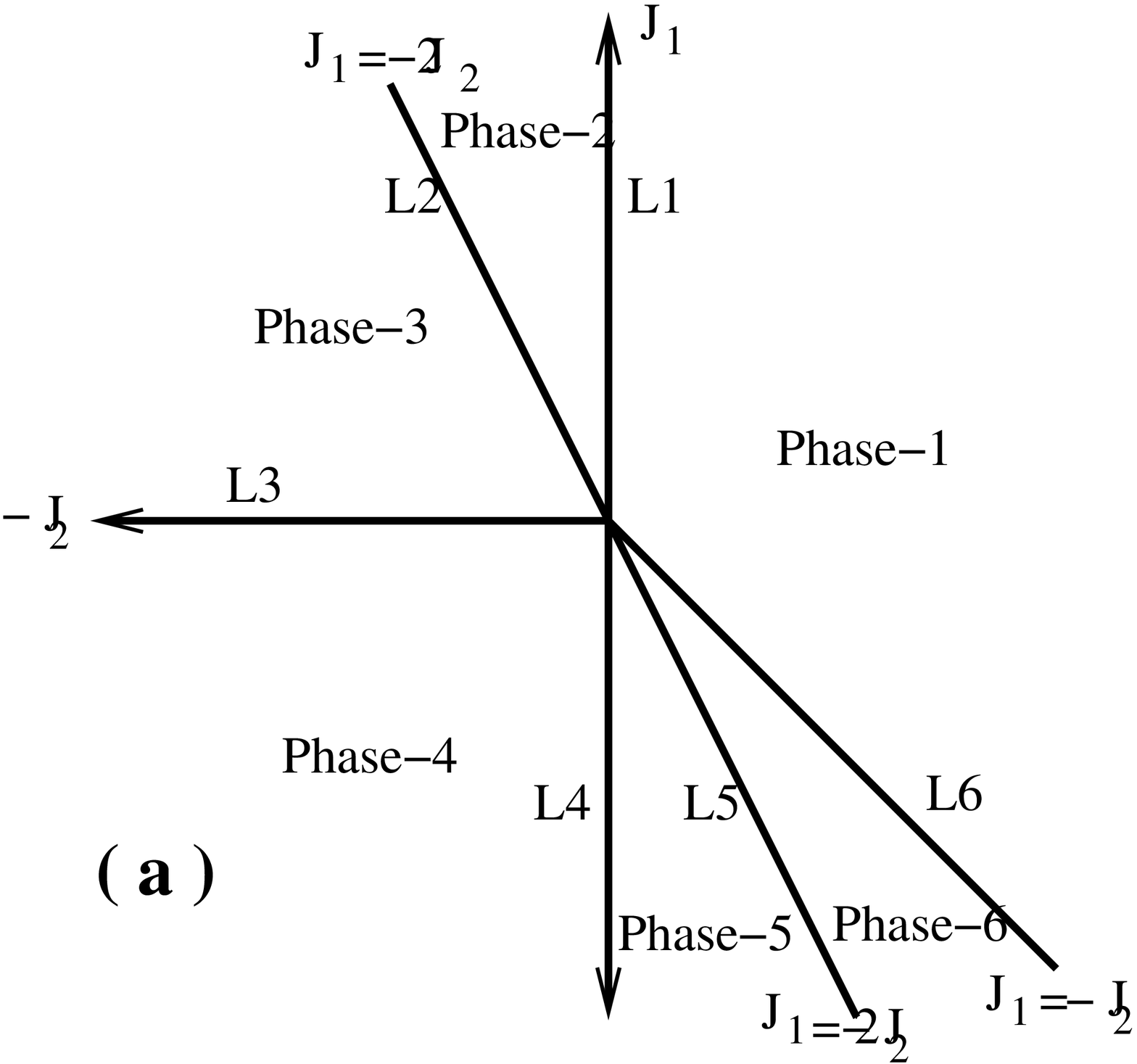}
\hspace{0.3cm}\includegraphics[scale=0.3]{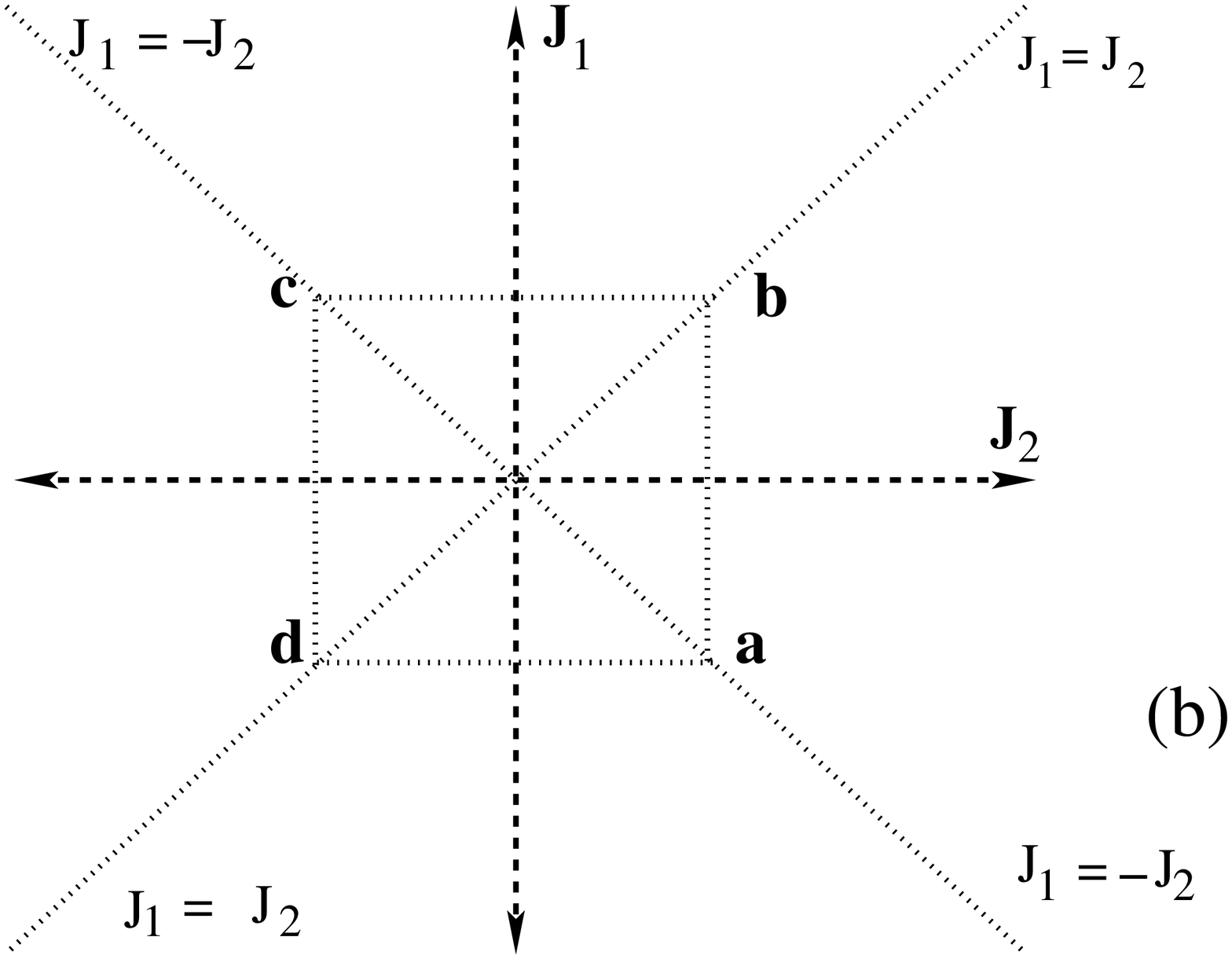}
\caption{\label{phasdiag}(a) $J_1$-$J_2$ phase diagram of the orbital structures in simple cubic lattice according to the new exchange interaction types $J_1$ and $J_2$ shown in Fig. \ref{Jtype1}: On the abscissa there is $J_2$, while the ordinate represents $J_1$. (b) The $J_1$-$J_2$ phase diagram, to simplify its study, is divided to region-1 from the line of the point $a=(J_1,J_2)=(-1,1)$ to the line of the point $b=(1,1)$, region-2 from $b$ to $c=(1,-1)$, region-3 from $c$ to $d=(-1,-1)$ and region-4 from $d$ to $a$. For more details see the text.}
\end{figure}
\section{Results and Discussion for the Regional Phases}
The simulations show that there is a different and unique order parameter for the six regions as shown in $J_1$-$J_2$ phase diagram in Fig. \ref{phasdiag}a. We describe each of these phases in turn. The ground state energy and transition temperature are obtained by both simulations and analytic reasoning along the four lines, $a\to b$, $b\to c$, $c\to d$ and $d\to a$ as presented below.
\vspace{0.1cm}
\begin{figure}[h!]
\centering
\includegraphics[scale=1]{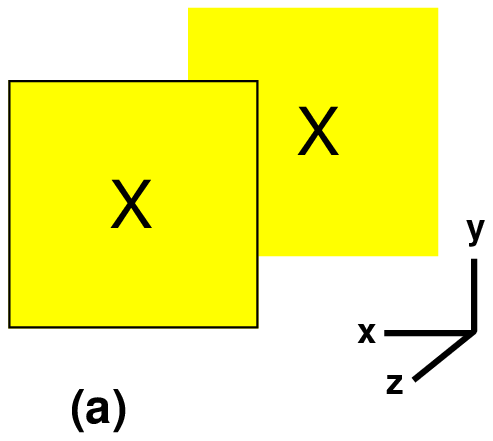}
\hspace{1cm}\includegraphics[scale=1]{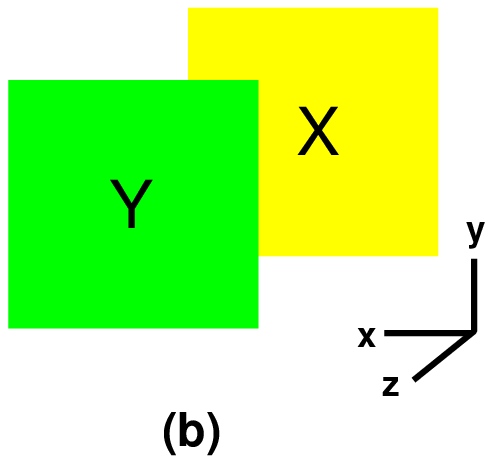}
\vspace{0.5cm}\hspace{1cm}\includegraphics[scale=1]{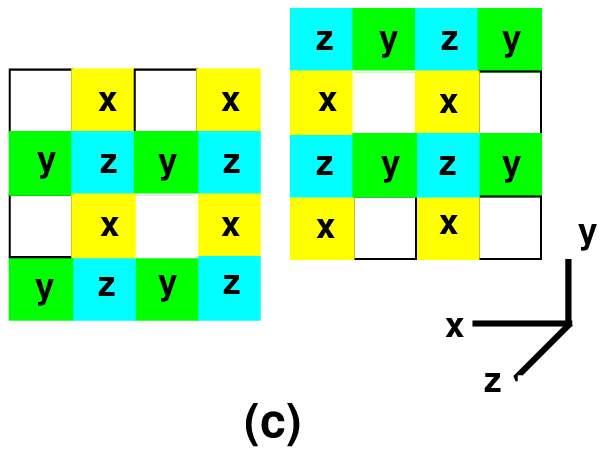}
\hspace{1cm}\hspace{1cm}\includegraphics[scale=1]{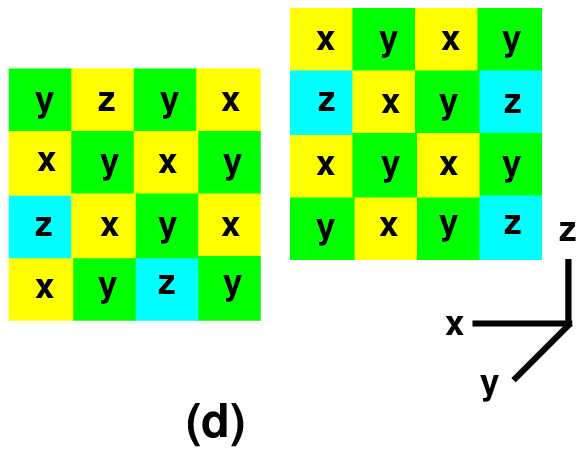}
\vspace{0.5cm}\hspace{1cm}\includegraphics[scale=1]{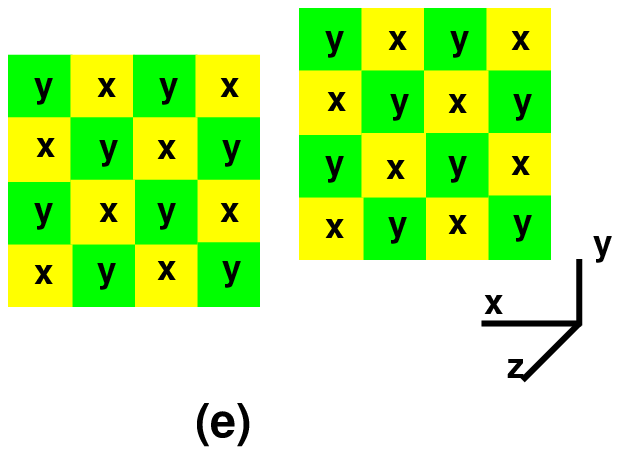}
\hspace{1cm}\includegraphics[scale=1]{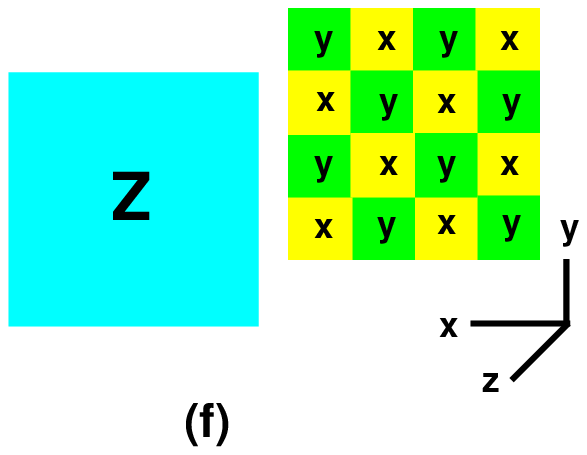}
\caption{\label{configurations}Schematic illustration for the ground state configurations for the regional phases obtained by the MC simulation of our model, (a) Phase-1, entirely ordered in each direction, (b) phase-2 is inserted between $L_1$ and $L_2$ in the $J_1$-$J_2$ phase diagram. The $x$- and $y$-layers are alternating in the $\hat{z}$ direction, (c) phase-3, in the $x$-$y$ plane, the $z$-state is ordering antiferromagnetically with $y$-state in $\hat{x}$ direction and with $x$-state in $\hat{y}$ direction, (d) phase-4 the well known AF Potts ground state, (e) phase-5 corresponds to the orbital ordering in manganites, (f) phase-6 is alternate ferromagnetic sheets and checkerboard layers. In all cases only one configuration is shown.}
\end{figure}

\subsection{\bf Phase-1}
\begin{figure}[h!]
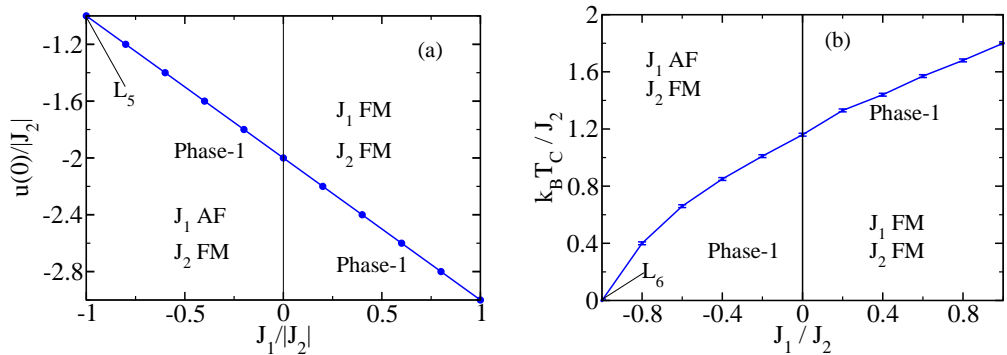

\centering
\includegraphics[scale=0.25]{A4_J1-U.eps}
\hspace{0.5cm}\includegraphics[scale=0.25]{AA4_J1-Tc.eps}
\caption{\label{area4}$J_1$-dependence of (a) the ground state energy, $u(0)$, and (b) transition temperature, $T_C$, from $(J_1,J_2)=(-1,1)$ to $(1,1)$ along the line a-b.}
\end{figure}

This is the well studied ferromagnetic phase that has an entirely ordered ground state configuration, Fig. \ref{configurations}a. This phase has a three dimensional order parameter corresponding to ordering of $x$, $y$ or $z$ orbits. The simulations show that the region of stability extends from $L_6$ where $J_2>0$ and $J_1>-J_2$ to $L_1$ where $J_1>0$ and $J_2>0$ (see figure \ref{phasdiag}a). The analytic expression for the ground state energy is given by,
\begin{eqnarray}
\label{uphase1}
u_1(0)=-J_1-2J_2,
\end{eqnarray}
when $J_1>0$, there is no frustration in this phase. When $J_1<0$ the optimal energy would be $u_{opt}=-2J_2$ and so in this case the ground state energy in the ferromagnetic phase is not optimal. The stabilization energy of this phase is obtained as, $\Delta u_1=u_1(\infty)-u(0)=\frac{2}{3}(J_1+2J_2)$. Since the ground state is completely ordered, as shown in Fig. \ref{configurations}a, we expect that the value of the ground state entropy, $s_1(0)$, vanishes and this was confirmed by our simulations with an error equal to $\pm0.005$. In Figs. \ref{area4}a and \ref{area1}a we show the excellent agreement between the simulated (solid circles) and analytic (solid line) values for the ground state energy, $u_1(0)$, as a function of $|J_1|/J_2$ along the line a-b and $|J_2|/J_1$ along the line b-c, where the lines were defined in Fig. \ref{phasdiag}b.
\vspace{0.1cm}

	Figures. \ref{area4}b and \ref{area1}b show the transition temperature, $T_C$, as a function of $J_1/J_2$ along the line a-b and $J_2/J_1$ along the line b-c in Fig. \ref{phasdiag}b. We note that the FM Potts model for $J_1=J_2$ is well studied and that value of $T_C$ we obtained from the simulation, $T_C=1.8\pm0.02$ $k_B/J_1$ is in agreement with the value obtained in Ref. \cite{Knak79}. This point is marked by a circle on Figs. \ref{area4}b and \ref{area1}b. We shall return to the subject of the specific heat in phase-1 in the region where $J_1<0$ in the section dealing with $L_6$.
\vspace{0.1cm}
\begin{figure}[h!]
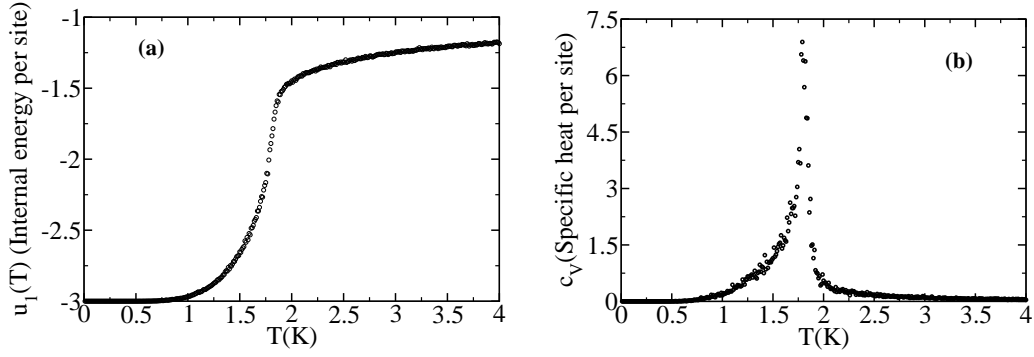

\centering
\includegraphics[scale=0.25]{fmpmu.eps}
\hspace{0.5cm}\includegraphics[scale=0.25]{fmpmcv.eps}
\caption{\label{u1}Monte Carlo simulation of the $T$-dependence of (a) ground state energy, $u_1(0)$, (b) the specific heat showing the transition temperature, $T_C=1.8 K$ for $J_1=J_2=1$.}
\end{figure}

	Summarising, we have reported known results for the ferromagnetic phase where it is clear that our simulation confirmed the well known result that the phase transition of  FM Potts model is first order, as seen in the plot of the temperature variation of the internal energy, $u_1(T)$, and specific heat, $c_V(T)$, in Fig. \ref{u1}, in agreement with other simulation studies and with the prediction of the $\epsilon$ expansion\cite{Rudnick75} but in contrast to the result of the position-space renormalization-group calculations\cite{Burkhardt76}.
\subsection{\bf Phase-2}
\begin{figure}[h!]
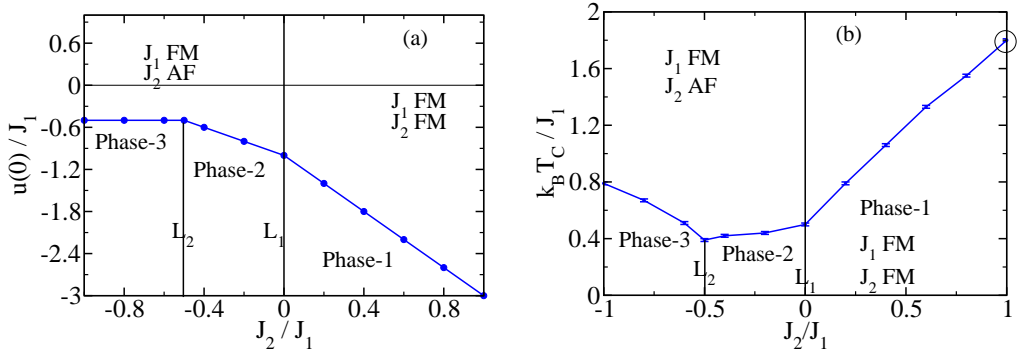

\centering
\includegraphics[scale=0.25]{A1_J2-U.eps}
\hspace{0.5cm}\includegraphics[scale=0.25]{AA1_J2-Tc.eps}
\caption{\label{area1}$J_2$-dependence of (a) the ground state energy, $u(0)$, and (b) transition temperature, $T_C$, from $(J_1,J_2)=(1,1)$ to $(1,-1)$ along the line b-c.}
\end{figure}

In phase-2, ordered ferromagnetic layers (OFM) exist between $L_1$ and $L_2$ for $-J_1/2<J_2<0$. In this region $J_2$ has become weakly AF and $J_1$ is still strongly ferromagnetic (see figure \ref{phasdiag}a). The ground state configuration consists of alternating FM layers, for example X,Y,X,Y,...., along the $\hat{z}$ direction as seen in Fig. \ref{configurations}b. In this phase, in an $X$ layer, an $x$ site has four nearest neighbours which are also $x$-state, two along the $\hat{x}$ direction and two along the $\hat{y}$ direction, and zero energy between any two neighbouring planes. In this case the ground state degeneracy is 6 as there are three ways to choose the normals to the plane and then a further factor of two to form the antiferromagnetic arrangement. The ground state energy, $u_2(0)$, is,
\begin{equation}
u_2(0)=-J_1-J_2.
\label{u2}
\end{equation}
The energy required to get this phase ordered is, $\Delta u_2 = - \frac{1}{3} (J_1+2J_2) + (J_1+J_2) = \frac{1}{3} (2J_1+J_2)$. This phase has long range order, hence, its ground state entropy, $s_2(0)$, is $=0.0\pm 0.006$ as is confirmed by simulation. Since $J_2<0$ this energy is not optimal. The lowest energy would be $-J_1$ but this is frustrated. Because the specific heat as function of $T$ obtained from the simulation for phase-2 is as sharp as that for phase-1 whose transition is well known, we believe that phase-2 has a first order transition. Figs. $\ref{area1}a$ and $\ref{area1}b$ show $u_2(0)$ and $T_C$ for this phase. The line phase $L_1$ occurs when $u_1(0)=u_2(0)$ where $J_1>0$ and $J_2=0$ at this line. The variation of the transition temperature with $J_2$ is also continuous at $J_2=0$.
\vspace{0.1cm}
\subsection{\bf Phase-3}
This phase is located when we move from $L_2$ to $L_3$ where $0<J_1<-J_2/2$ (see figure \ref{phasdiag}a). This occurs as $J_2$ becomes more strongly antiferromagnetic, and it has more complicated ordering. For clarity we showed in Fig. \ref{configurations}c only the $3/4$ of sites that have long range order. These form a 3d network of antiferromagnetic chains whose energy is zero. The sites left blank have two $x$ states nearest neighbours in the $\hat{x}$ direction, two $y$ states nearest neighbours in the $\hat{y}$ direction and two $z$ states in the $\hat{z}$ direction. The energy of a state on the blank site is exactly $-2J_1$ whichever state occupies this site. We call this phase the `Cage' phase because the energy comes from the sites left blank, as seen in Fig. \ref{configurations}c, that are in a cage. However, only $1/4$ of the sites are blanks. So the ground state energy, $u_3(0)$, for phase-3 is,
\begin{equation}
u_3(0)=\frac{1}{4}(-2J_1)=-\frac{1}{2}J_1.
\label{u3}
\end{equation}
Because the occupation of 1/4 of the sites may be chosen randomly, the analytic expression for the entropy is,
\begin{equation}
s_3(0)/k_B=\frac{1}{4}log_e3\simeq 0.27465.
\label{s3}
\label{phase3s}
\end{equation}
The energy needed to order this phase is, $\Delta u_3=- \frac{1}{3} (J_1+2J_2)+\frac{1}{2}J_1=\frac{1}{3}(\frac{1}{2}J_1-2J_2)$.

\begin{figure}[h!]
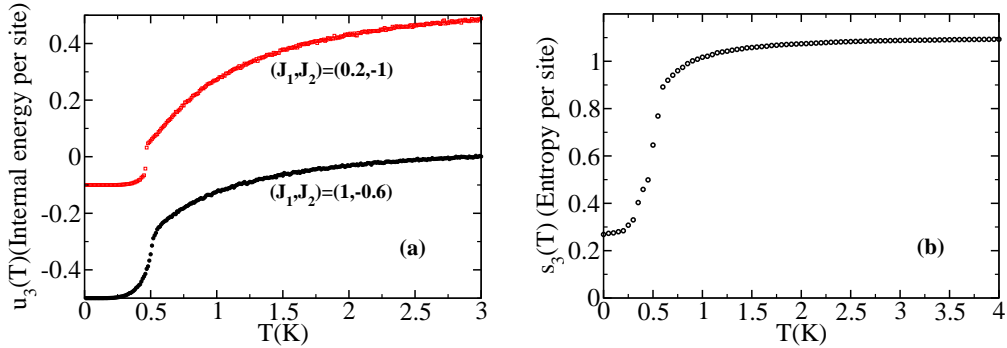

\centering
\includegraphics[scale=0.25]{cageu.eps}
\hspace{0.5cm}\includegraphics[scale=0.25]{cages.eps}
\caption{\label{cage}MC simulations of the $T$-dependence of (a) the internal energy, $u_3(T)$, at the beginning of the phase-3 region at $J_2/J_1=-0.6$ and the end of the region at $J_2/J_1=-5$ (b) entropy, $s_3(T)$, at $J_2/J_1=-0.6$.}
\end{figure}

	Again, we have a frustrated phase because of the competition between the strong AF $J_2$ and FM $J_1$. If it were possible to arrange the Potts model so that each one had an interaction energy $-J_1$ with two nearest neighbours and zero interaction with the nearest neighbours in plane, the energy per site would be $-J_1$. Figs. \ref{area1}a and \ref{area2}a show that there is excellent agreement between the simulated and analytic results of the ground energy, $u(0)$, as a function $J_2/J_1$ along the line b-c and $J_1/|J_2|$ along the line c-d. In this case, the ground state entropy obtained by the simulation, at the point $J_2/J_1=-0.6$, is $0.27 \pm 0.005$ (see Fig. \ref{cage}b), and this value is in fair agreement with the value which is predicted analytically Eq. (\ref{s3}).
\vspace{0.1cm}

	Additionally, when the ground state energy of phase-2 is equal to that for phase-3, $u_2(0)=u_3(0)$, the line phase $L_2$ is obtained. From Eqs. (\ref{u2}) and (\ref{u3}), we get $J_2=-\frac{1}{2}J_1$ which agrees with the phase boundary $L_2$ obtained from the simulation.

	This phase has a first order transition as $J_1\to 0$, as seen from the plot of $u_3(T)$ with $T(K)$ in Fig. \ref{cage}a. The simulated specific heat is similar for this phase to that for phase-1. The simulation is able to give an accurate value of the ground state entropy to the analytic value in the beginning of the phase-3 region but not at the end of this phase where the transition temperature is approaching zero.

	It is shown that $T_C$, obtained from the simulation, as function of $J_2$ for phase-3 region in Fig. \ref{area1}b, increases dramatically and $T_C$ as function of $J_1$, Fig. \ref{area2}b, decreases promptly to zero at $L_3$ where there is no a phase transition. The value of $T_C$ is obtained from the simulation and is shown in Fig. \ref{area1}b for $-1\leq J_2/J_1 \leq -0.5 $ and in Fig. \ref{area2}b for $-1\leq J_2/J_1\leq 0$. The transition temperature falls to zero as $J_1$ approaches zero at $L_3$.

\subsection{\bf Phase-4}
This is the 'AF' phase which is well known and has been studied extensively \cite{Banavar82,Wang90,Grest81}. This phase occupies the whole quadrant where both $J_1$ and $J_2$ are AF and it is located between $L_3$ at the point (0,-1) and $L_4$ at the point (-1,0), see $J_1-J_2$ phase diagram in Fig. \ref{phasdiag}a. Each site in this phase is surrounded with different orbitals (see figure \ref{configurations}d), the ground state energy through the whole phase is
\begin{equation}
u_4(0)=0,
\label{u4}
\end{equation}
where the stabilization energy of this phase is, $\Delta u_4=-\frac{1}{3} (J_1+2J_2)+0=-\frac{1}{3} (J_1+2J_2)$.
\vspace{0.2cm}
\begin{figure}[h!]
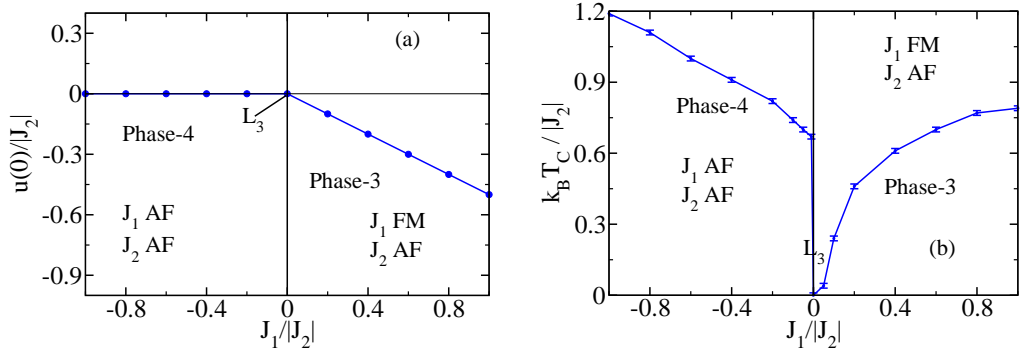

\centering
\includegraphics[scale=0.25]{A2_J1-U.eps}
\hspace{0.5cm}\includegraphics[scale=0.25]{AA2_J1-Tc.eps}
\caption{\label{area2}$J_1$-dependence of (a) the ground state energy, $u(0)$, and (b) transition temperature, $T_C$, at $1>J_1>-1$ and $J_2=-1$.}
\end{figure}

	The order is understood if the lattice is divided into two sublattices, where one of the three states is on the first sublattice and the other two states are distributed randomly on the second sublattice\cite{GGrest81}. This leads to a ground state entropy per site of $\frac{1}{2}k_Blog_e2$. But, sometimes, at $T=0$, states of the lattice are on the wrong sublattice if the surrounding neighbours permit it\cite{Banavar82}. Accurate Monte Carlo estimates including finite size scaling have evaluated \cite{Wang90}, $s_4(0)=0.3673k_B$. This is higher than the estimated, $s_4(0)=\frac{1}{2}k_Blog_e2$, by an amount $=0.0207k_B$. We can quantify the argument of Banavar {\it et al} (1982)\cite{Banavar82} as follow. The probability that a site on the ordered sublattice has six identical neighbours is $2(\frac{1}{2})^6$. In this case the site on the ordered sublattice can take one of two values. This gives an analysis for the entropy,
\begin{equation}
s_4(0)=\frac{k_B}{2}log_e2+\frac{k_B}{32}log_e2\simeq 0.3682k_B.
\end{equation}
This is in good agreement with Wang {\it et al}\cite{Wang90}. There are six ways of defining the ordered part of the ground state coming from two ways of defining the sublattice and three choices of the orbit that orders. We find the same ground state configurations throughout this phase and the value of the ground state entropy also takes the same value throughout this phase.
\vspace{0.2cm}
\begin{figure}[h!]
\centering
\includegraphics[scale=0.3]{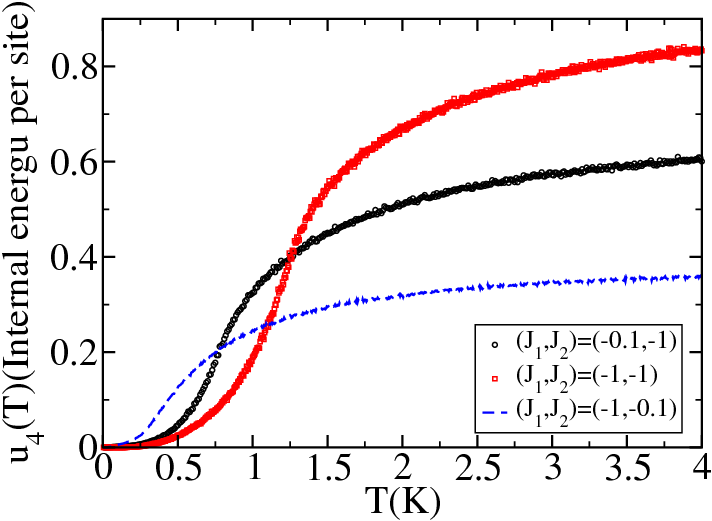}
\caption{\label{afpmu}The internal energy, $u_4(T)$, vs $T(K)$ at $(J_1,J_2)=(-0.1,-1),(-1,-1)$ and (-1,-0.1).}
\end{figure}

	There is good agreement between the simulated results of the ground state energy and the analytic results as function of $J_1/|J_2|$ along the line c-d and $J_2/|J_1|$ along the line d-e, as seen in Figs. \ref{area2}a and \ref{area3}a. Fig. \ref{afpmu} seems to imply that Phase-4 has a continuous transition but the possibility of very weak first order transitions cannot be excluded\cite{Banavar82} because the simulation was done on a small cluster.
\subsection{\bf Phases -5 and -6}
The region where $J_1$ is negative and $0<J_2<1$ is very interesting because it is divided into two related phases. The phases have exactly the same ground state energy but very different configurations. The ground state configurations are shown in Figs. \ref{configurations}e and \ref{configurations}f.

	In phase-5 we have an $x,y$ checkerboard pattern in the $x-y$ plane and the planes are stacked so that the $x$ and $y$ states are in ordered chains up the $\hat{z}$ axis. This is the pattern of orbits seen in $LaMnO_3$ ( see Fig. \ref{OO}) so we call it LMO phase. The contribution to the energy comes from these ferromagnetic chains so the ground state energy is given by $u_5(0)=-J_2$.

\begin{figure}[h!]
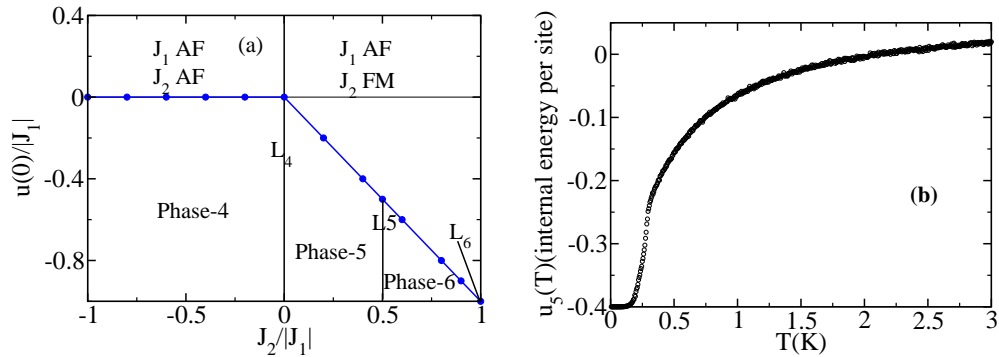

\centering
\includegraphics[scale=0.25]{A3_J2-U.eps}
\hspace{0.5cm}\includegraphics[scale=0.25]{LMOu.eps}
\caption{\label{area3}(a) $J_2$-dependence of the ground state energy, $u(0)$, at $-1<J_2<1$ and $J_1=-1$, (b) $T$-dependence of the internal energy, $u_5(T)$, at the point $J_2/J_1=-0.4$ in phase-5 region.}
\end{figure}

	In phase-6 we have an alternation up the $\hat{z}$ direction of an $x$-$y$ checkerboard layer with a layer that is occupied by $z$ orbits. We call this the FM-CB phase. The contribution to the ground state energy comes from the ferromagnetic layers. Each $z$ orbit has four $z$ nearest neighbours in the $\hat{x}$ and $\hat{y}$ directions and so the energy per site of the phase is $-2J_2$. However, only half of the planes are ferromagnetic so the total energy per site is half the energy from the planes which is the same as phase-5. The analytic ground state energy per site in phase-5 and phase-6  and the stabilization energies are given below,
\vspace{0.2cm}
\begin{equation}
u_5(0)=u_6(0)=-J_2.
\label{u5,6}
\end{equation}
\begin{equation}
\Delta u_5(0)=\Delta u_6(0)=-\frac{1}{3}(J_1-J_2).
\end{equation}
Both of these are frustrated in the sense that the lowest possible ground state energy of $-2J_2$ is not accessible. The Monte Carlo calculations give the ground state energy in agreement with Eq. (\ref{u5,6}) as shown in Fig. \ref{area3}a.

	The region occupied by phase-5 needs either more Monte Carlo steps or a larger lattice size in order to reach a pure LMO ground state. If the system does not come into equilibrium then a mixed ground state occurs that includes some of phase-6.
\vspace{0.2cm}

\begin{figure}[h!]
\centering
\includegraphics[scale=1]{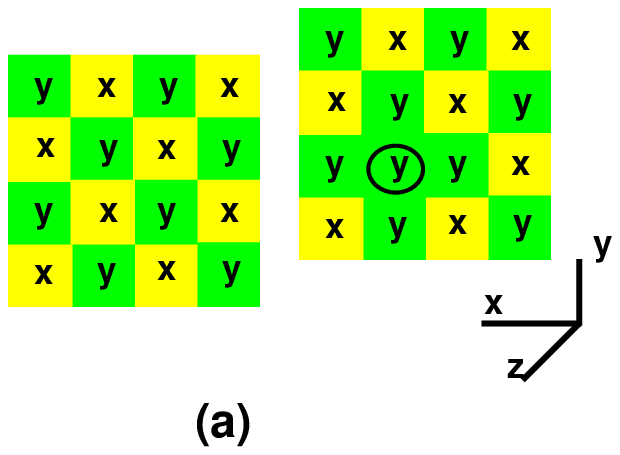}
\hspace{1cm}\includegraphics[scale=1]{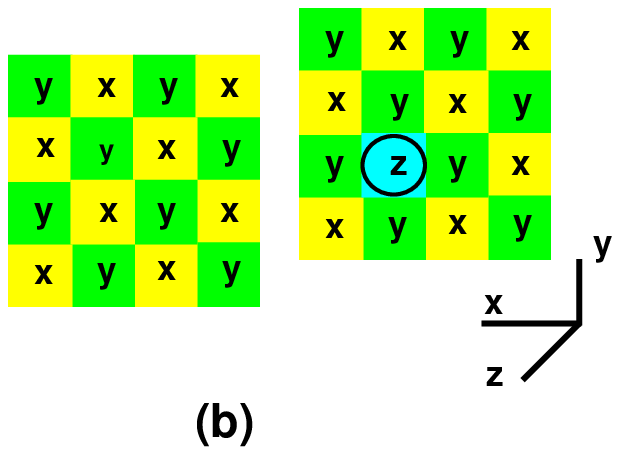}
\caption{\label{CBexcite} A single defect in the LMO phase. The site marked with a circle has flipped from $x$ (a) to $y$ and (b) to $z$.}
\end{figure}

\begin{figure}[h!]
\centering
\includegraphics[scale=1]{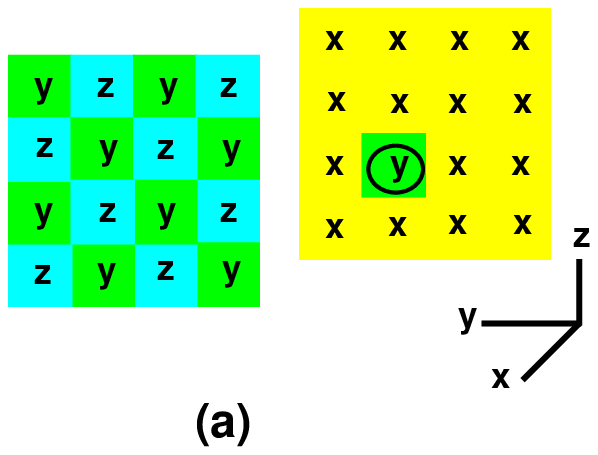}
\hspace{1cm}\includegraphics[scale=1]{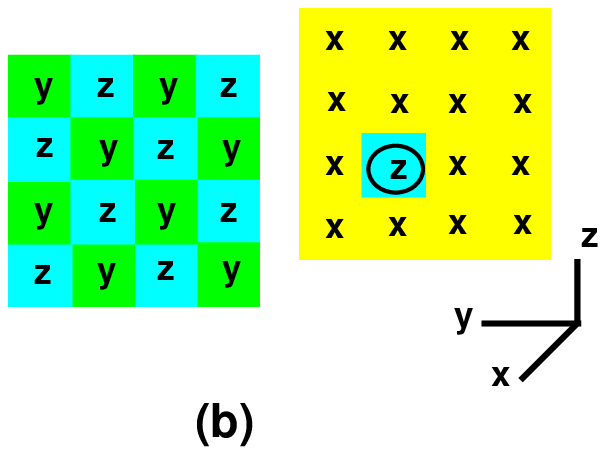}
\hspace{1cm}\includegraphics[scale=1]{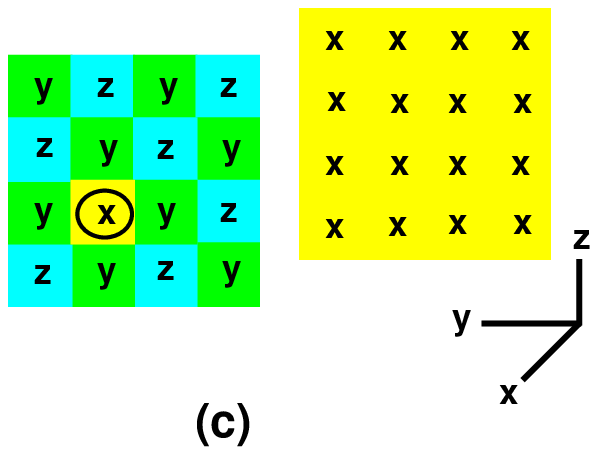}
\hspace{1cm}\includegraphics[scale=1]{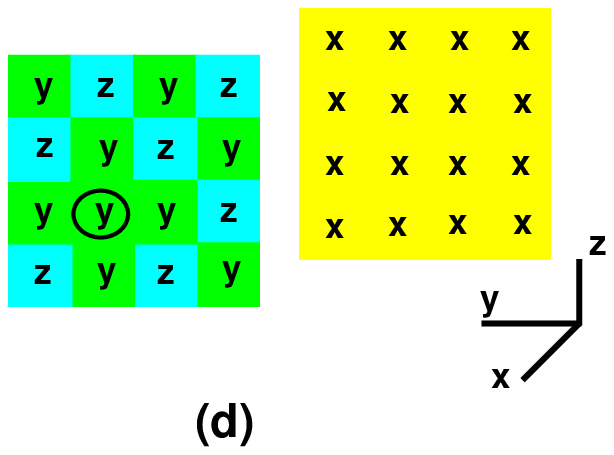}
\caption{\label{FM-CBexcite} A single defect in the FM-CB phase. The site marked with a circle in FM layer has flipped from $x$ (a) to $y$ and (b) to $z$. And, the site marked with circle in CB layer has flipped from $z$ (c) to $x$ and (d) to $y$.}
\end{figure}

	Because the ground state energies of phase-5 and phase-6 are equal we need a more sophisticated argument to obtain the phase diagram. We compare the free energy of the two phases in the limit as the temperature approaches zero. This is an example of a phase stabilized by disorder, 'order by disorder',\cite{Chalker92,Ritchy93} which has been used to study frustrated Heisenberg models. We note that we are considering a broken discrete symmetry compared with the continuous symmetry problems discussed earlier. At $T\neq 0$, we calculate the free energy for each phase individually. We write $F(T)=-J_2+\Delta F(T)$ and evaluate $\Delta F$.
\begin{equation}
\Delta F=-k_BTlog_eZ.
\end{equation}
The partition function, $Z$, is evaluated from the energies $\epsilon _i$ of excitations away from the ground state where $\beta = 1/k_BT$.
\begin{equation}
Z=1+\sum_{i}e^{-\beta\epsilon _i}.
\end{equation}
In the case of phase-5 the LMO phase, we can flip an $x$ site to either $y$ or $z$ as shown in Figs. \ref{CBexcite}a and \ref{CBexcite}b. The changes in energy associated with the ringed site in cases (a) and (b) are given,
\begin{equation}
\Delta \epsilon_a=-J_1-J_2+J_2=-J_1,
\end{equation}
\begin{equation}
\Delta \epsilon_b=+J_2.
\end{equation}
\vspace{0.2cm}
In both cases there is a factor of $J_2$ coming from the change in energy per site associated with the bonds in the $\hat{z}$ direction. We note that $J_1$ is negative in this region, so, both energies $\Delta \epsilon_a$ and  $\Delta \epsilon_b$ are positive. The energies of flipping a $y$ site to either $x$ or $z$ are the same leading to the following expression for $Z_5$,
\begin{equation}
Z_5=1+2e^{J_1\beta}+2e^{-J_2\beta}.
\end{equation}
This leads to an expression for $\Delta F_5$ in the low temperature limit,
\begin{eqnarray}
\Delta F_5&=&-Nk_BTlog_e[1+2e^{J_1\beta}+2e^{-J_2\beta}]       \nonumber \\
&\cong &-2Nk_BT[e^{J_1\beta}+e^{-J_2\beta}].
\end{eqnarray}
We now consider the FM-CB layer phase. In this case we have four possibilities. We can flip a site in the ferromagnetic plane to either $x$ or $y$ as shown in Figs. \ref{FM-CBexcite}a and (\ref{FM-CBexcite}b) or we can flip a site in the checkerboard plane as shown in Figs. \ref{FM-CBexcite}c and \ref{FM-CBexcite}d.
\vspace{0.2cm}
The change in the energy per site associated with the ringed site in cases (a) and (b) is
\begin{eqnarray}
\Delta \epsilon_a &=& 0-(-2J_2)=2J_2            \nonumber \\
\Delta \epsilon_b &=& -J_2-(-2J_2)=J_2,
\end{eqnarray}
where, in the FM layers, the ground state energy per site is $u_{a,b}=-2J_2$. In cases (c) and (d) $\Delta \epsilon$ is
\begin{eqnarray}
\Delta \epsilon_c&=&-J_1-0=-J_1                         \nonumber \\
\Delta \epsilon_d&=&-(J_1+J_2)-0=-(J_1+J_2),
\end{eqnarray}
where the ground state in CB layer case is zero. We note that in phase-6, $J_1<0$ and $J_2<|J_1|$, so, all the energies $\Delta \epsilon_a,.., \Delta \epsilon_d$ are positive. The partition function for phase-6, $Z_6$, is
\begin{eqnarray}
Z_6=1+e^{-2J_2\beta}+e^{-J_2\beta}+e^{J_1\beta}+e^{(J_1+J_2)\beta}.
\end{eqnarray}
The free energy $\Delta F_6$ is
\begin{eqnarray}
\Delta F_6&=&-Nk_BTlog_e[1+e^{-2J_2\beta}+e^{-J_2\beta}+e^{J_1\beta}+e^{(J_1+J_2)\beta}] \nonumber \\
&\cong &-Nk_BT[e^{-2J_2\beta}+e^{-J_2\beta}+e^{J_1\beta}+e^{(J_1+J_2)\beta}].
\end{eqnarray}
The stable phase will be the one with the lower free energy.
\begin{eqnarray}
\Delta F_5-\Delta F_6&=&-Nk_BT[2e^{J_1\beta}+2e^{-J_2\beta}-e^{-2J_2\beta}-e^{-J_2\beta}-e^{J_1\beta}- e^{(J_1+J_2) \beta} ]        \nonumber \\
&=& Nk_BT [(e^{J_2\beta}-1) (e^{J_1\beta} - e^{-2J_2\beta})].
\end{eqnarray}

	Phase-5 and phase-6 exist in a region where $J_2>0$ and $J_1<0$. The condition for phase-6 to be stable is that $(\Delta F_5-\Delta F_6)>0$. This is given by $(2J_2+J_1)>0$. Thus, we find that phase-5 is stable for $J_1<0$ and $0<J_2<-J_1/2$. The boundary between phases-5 and -6 comes at $J_1=-2J_2$ and phase-6 is stable for $-2J_2<J_1<-J_2$. The ground state configuration obtained by the MC simulation confirms this analytic result, see Fig. \ref{configurations}e for phase-5 and Fig. \ref{configurations}f for phase-6. The ground state of these phases are expected to have zero entropy and this is also confirmed by the simulation with an error equal to $\pm0.02$.

\begin{figure}[h!]
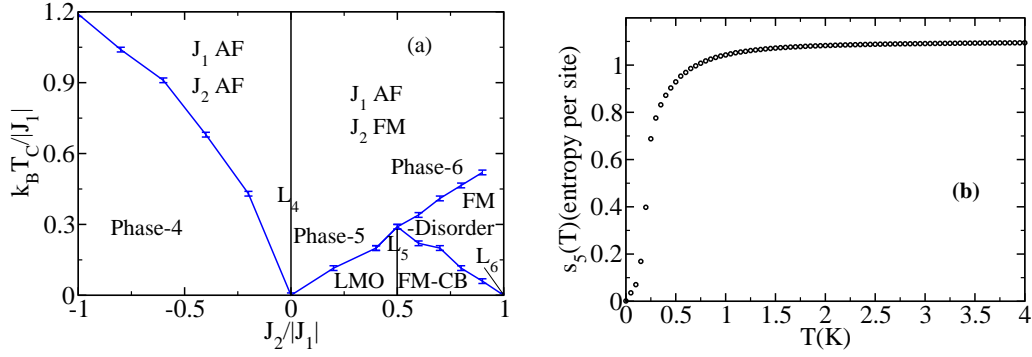

\centering
\includegraphics[scale=0.25]{AA3_J2-Tc.eps}
\hspace{0.5cm}\includegraphics[scale=0.25]{LMOs.eps}
\caption{\label{LMOus}(a) $J_2$-dependence of the transition temperature, $T_C$, from $(J_1,J_2)=(-1,-1)$ to $(-1,1)$, (b) $T$-dependence of the entropy, $s_5(T)$, at the point  $(J_1,J_2)=(-1,0.4)$ in phase-5 region.}
\end{figure}

	It is clear from the Fig. \ref{area3}a that the analytic ground state energy, $u(0)$, as a function of $J_2/|J_1|$ agrees very well with the simulated results. In addition, it decreases continuously and steadily from $L_4$ to $L_6$ with increasing $J_2$. One can easily notice that the condition for $L_4$ which is $u_4=u_5$ is for $J_2=0$ and when $u_5=u_6$ the line phase $L_5$ occurs at $(J_1,J_2) = (-1,0.5)$. When $u_6 = u_1$ the line phase $L_6$, $J_2=-J_1$, is found to separate phase-6 and phase-1. However, the transition temperature, $T_C$, for phase-5 increases slightly, as seen in Fig. \ref{LMOus}a, from zero at $L_4$ to $0.29\pm 0.01$ at $L_5$, with increasing $J_2$. Fig. \ref{area3}b shows the T-dependence of the internal energy, $u_5(T)$, and Fig. \ref{LMOus}b shows the entropy, $s_5(T)$, at the point $(J_1,J_2)=(-1,0.4)$. We compare the behaviour of specific heat for phase-5 and the upper transition for phase-6 with that for phase-1 to deduce that both have a first order transition.
\vspace{0.2cm}

\begin{figure}[h!]
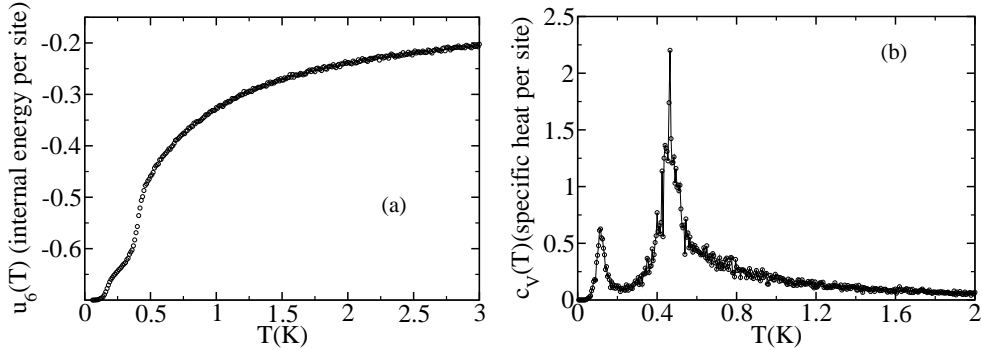

\centering
\includegraphics[scale=0.25]{FM-CBu.eps}
\hspace{0.2cm}\includegraphics[scale=0.25]{FM-CBcv.eps}
\caption{\label{FM-CB}$T$-dependence of (a) the internal energy, $u_6(T)$, and (b) the specific heat, $c_V(T)$, at the point  $(J_1,J_2)=(-1,0.7)$ in phase-6 region.}
\end{figure}

	The phase diagram in phase-6 region is highly unusual for a Potts model because there are two district ordering temperatures. These are seen clearly in the specific heat as shown in Fig. \ref{FM-CB}b. We looked at configurations from the Monte Carlo simulations in the intermediate phase in phase-6 region. These had long range order developing in alternate ferromagnetic planes, say, in orbital $z$ as shown in  Fig. \ref{configurations}f and the intermediate layers were disordered but contained predominantly orbitals $x$ and $y$.

	This may be understood analytically. The ground state energy for the FM layers is $u_{FM}(0)=-2J_2$, the energy needed to order this layer is
\begin{eqnarray}
\Delta u_{FM}&=&u_{FM}(0)-u(\infty)          \nonumber \\
&=& -2J_2+\frac{1}{3}(J_1+2J_2) = \frac{1}{3}(J_1-4J_2),
\end{eqnarray}
The ground state energy for the CB layers only is zero, so, the energy required to these layers to be ordered is
\begin{eqnarray}
\Delta u_{CB}&=&u_{CB}(0)-u(\infty)          \nonumber \\
&=& 0+\frac{1}{3}(J_1+2J_2) = \frac{1}{3}(J_1+2J_2).
\end{eqnarray}
We check the difference between, $\Delta u_{CB}$ and $\Delta u_{FM}$ as follows,
\begin{eqnarray}
\Delta u_{CB}-\Delta u_{FM}=\frac{1}{3}(J_1+2J_2)-\frac{1}{3}(J_1-4J_2) = 2J_2.
\end{eqnarray}
Since $J_2>0$, $\Delta u_{CB} > \Delta u_{FM}$. This means that the FM layers order first at $T_{C1}$, before the CB layers order at $T_{C2}$, where $T_{C1}>T_{C2}$ as seen in Fig. \ref{LMOus}a.
\vspace{0.2cm}

	The unusual feature of this phase is that when the ferromagnetic planes are formed the orbits on the intermediate planes are free to order independently. The $x$ and $y$ orbitals on the sites on the intervening planes order antiferromagnetically at $T_{C2}$. We expected that this transition should belong to the class of two dimensional Ising models and should be second order. We believe that our simulations are in agreement with this conjecture because the peak in the simulated $T$-dependence of the specific heat for phase-6 at the lower transition temperature $(T_{C_2})$, see Fig \ref{FM-CB}b, in less pronounced than that for the first order transitions we have studied. The peak is similar to that for the antiferromagnetic phase (phase-4) which is obtained to be second order . There is no reason for the antiferromagnetic order parameter to be coherent up the $\hat{z}$ direction so this phase would show disorder scattering down to low temperatures. However, the entropy per site would vary as $k_BL^{-2}log_e2$ and would vanish in the thermodynamic limit. This is confirmed by the simulations.

\vspace{0.2cm}
\begin{figure}[h!]
\centering
\includegraphics[scale=0.75]{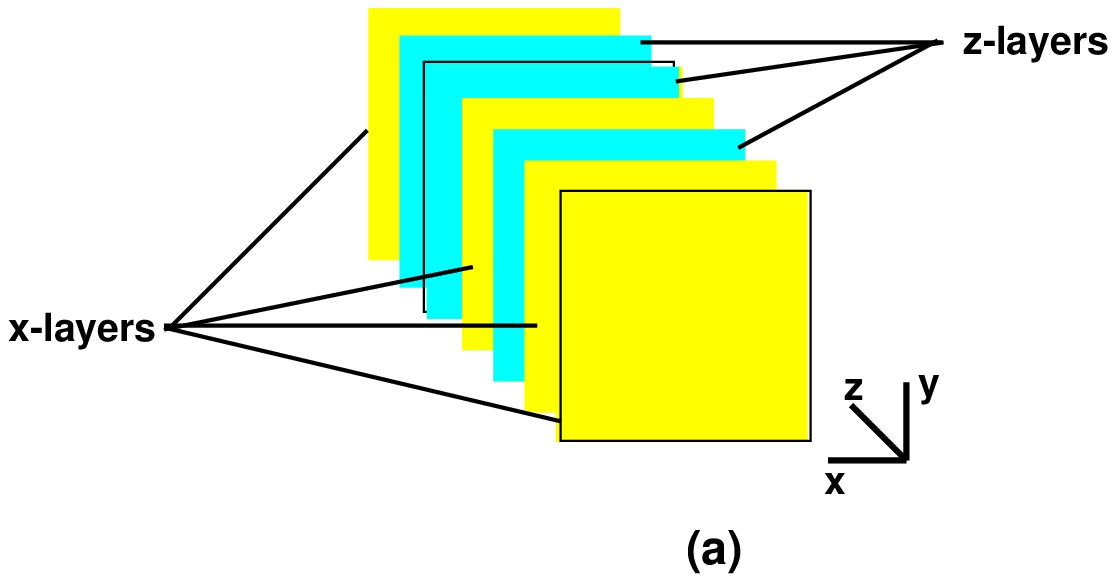}
\hspace{1cm}\includegraphics[scale=0.75]{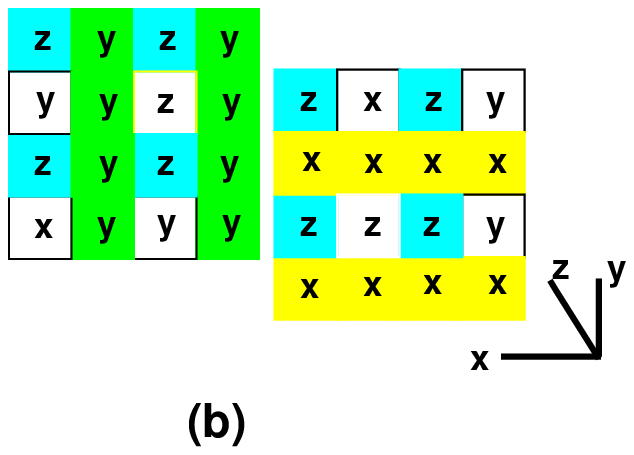}
\hspace{1cm}\includegraphics[scale=0.75]{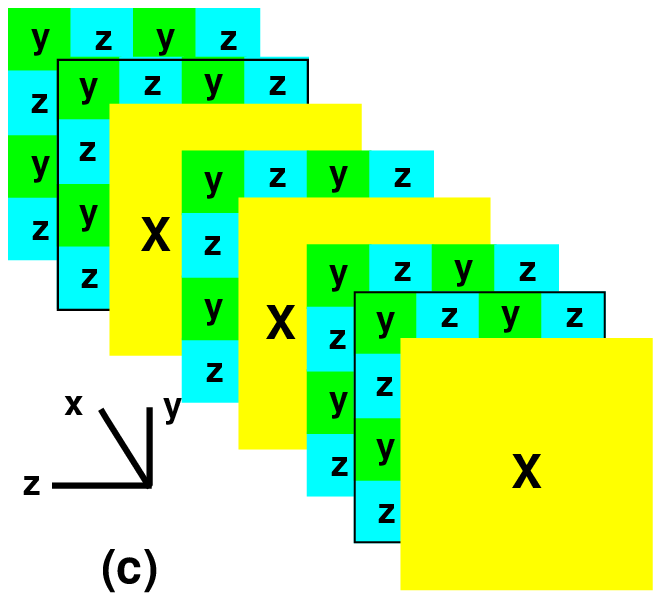}
\hspace{1cm}\includegraphics[scale=0.75]{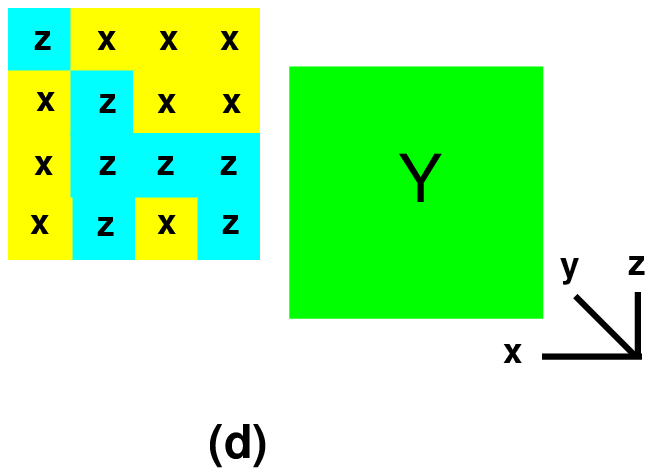}
\caption{\label{lines}Schematic shape for the line phases ground state ordering (a) 'RFM' layers phase along $L_1$ which is inserted between phase-1 and phase-2, (b) 'Wood Pile' phase along $L_2$ which is inserted between Phase-2 and phase-3, (c) FM and CB layers along $L_5$ alternate randomly with each other in $\hat{x}$ direction, so it is called 'RFMCB' phase. (d) 'FM-disorder' layers phase along $L_6$ which is located between phase-6 and phase-1.}
\end{figure}
\section{Results and Discussion for the Boundary Lines}
The thermodynamics of the boundary lines between two phases will differ from the phases on each side because in this case two different configurations can occur with the same ground state energy. Because of the extra allowed configurations boundary lines, the entropy will be greater or equal to that of both the adjoining phases in all cases. We investigate each boundary phase.\\

\subsection{\bf Phase $L_1$}
The phase on the line $L_1$ in the $J_1-J_2$ phase diagram is investigated. This line separates phase-1 ('FM' phase) and phase-2 ('OFM' Layers) where $J_1>0$ and $J_2=0$. This phase is a transition from ferromagnetically coupled layers to antiferromagnetically coupled layers. In this case we expect the planar order to be preserved on this line. Fig. \ref{lines}a shows that $L_1$ consists of different (x-layer and y-layer) FM layers alternating randomly with each other in $\hat{z}$ direction. We call this phase 'Random FM' Layers or 'RFM'. Each site has, in-plane, two nearest neighbours with $J_1$ interaction and two nearest neighbours with $J_2$ interaction. Then, the analytic formula representing the ground state energy per site for this phase for $J_2=0$ is
\begin{equation}
u_{L_1}(0)=-J_1.
\end{equation}
Its ordering energy is, $\Delta u_{L_1}=-J_1+\frac{1}{3}(J_1+2J_2)=\frac{2}{3}(J_2-J_1)$. Because the $L_1$ phase is completely ordered in two dimensions, its ground state entropy per site, $s_{L_1}(0)$, tends to vary as $L^{-2}log_e2$ and is zero in the thermodynamic limit.The simulated value is $0\pm0.008$.

	It is clear from Fig. \ref{area1}a that the simulated energy agrees with the analytic value for this line phase. The value of the transition temperature is continuous across $L_1$. As seen from the simulated plot of $T$-dependence of specific heat per site, $c_V(T)$, in Fig. \ref{Lcv}a the Monte Carlo simulations indicate that the transition at $L_1$ is similar to that in the ferromagnetic phase and hence is expected to be first order.

\subsection{\bf Phase $L_2$}
 This line at $J_2/J_1=-1/2$ separates the antiferromagnetic layer phase from the cage phase. On this line the AF effects of $J_2$ prevent the formation of the FM layers and the competition with the FM effect of $J_1$ yields FM chains perpendicular to AF chains as seen in Fig. \ref{lines}b.
\vspace{0.2cm}
\begin{figure}[h!]
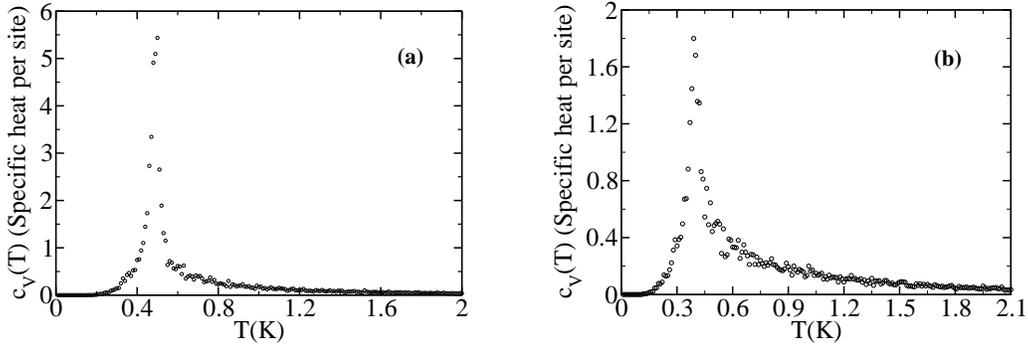

\centering
\includegraphics[scale=0.25]{L1cv.eps}
\hspace{1cm}\includegraphics[scale=0.25]{L2cv.eps}
\caption{\label{Lcv}$T$-dependence of specific heat, $c_V(T)$, (a) for the line phase $L_1$, (b) for the line phase $L_2$.}
\end{figure}

	This phase is obtained when $u_2(0)=u_3(0)$. We get this phase when $J_2=-\frac{1}{2}J_1$, any site could have two nearest neighbours with energy $-J_2$ per site and other two nearest neighbours with energy $-J_1$ per site. The total ground state energy is, $u_{L_2}(0)=-J_1-J_2$, but we know that $J_2=-\frac{1}{2}J_1$, hence,
\begin{equation}
u_{L_2}(0)=-\frac{1}{2}J_1,
\end{equation}
and its ordering energy is, $\Delta u_{L_2} = - \frac{1}{3}(J_1+2J_2)+\frac{1}{2}J_1 = -\frac{1}{3}(2J_2 - \frac{1}{2}J_1)$.

	This phase is called wood pile as there are ferromagnetic chains of $x$, $y$ and $z$ orbitals running along the $\hat{x}$, $\hat{y}$ and $\hat{z}$ directions respectively. This leaves one quarter of the sites disordered so the ground state entropy is given by $s_{L_2}(0) = \frac{k_B}{4} log_e3 \simeq 0.2746k_B$ which is confirmed by the simulations with an error equal to $\pm 0.001$.

	The simulated and analytic ground state energy $u_{L_2}(0)$ shown in Fig. \ref{area1}a agree. The transition temperature is continuous across $L_2$, Fig. \ref{area1}b. The plot of the specific heat shown in Fig. \ref{Lcv}b indicates a first order transition compared with the transition of phase-1.\\

\subsection{\bf Phase $L_3$}
This phase is at $J_1=0$ and $J_2<0$ and separates phase-3 and phase-4. The energy of this phase is found from this condition $u_3(0)=u_4(0)=0$. This is confirmed by simulation as shown in Fig. \ref{area2}a. There is no long range order hence no transition on this line which separates two phases having dissimilar long range order, phase-3 and phase-4. However, very short FM chains can be seen diagonally with short AF chains along the three axes. These different ways of ordering make the ground state entropy higher than that for long range ordered phases adjacent to this line.

	While, there is no analytic expression for the ground state energy and entropy, the simulated value for the ground state energy, $u_{L_3}(0)$ is equal to that for phase-4, $u_4(0)$, in Eq. (\ref{u4}) as seen in Fig. \ref{area2}a. The entropy value obtained by the simulation is, $s_{L_3}(0)=(0.423\pm0.003)k_B$. Fig. \ref{area2}b shows that there is discontinuity in the $J_1$-dependence of $T_C$ at $L_3$. In the phase-3 region, $T_C$ decreases with decreasing of $J_1/|J_2|$ and goes to zero at $L_3$, but suddenly jumps to a finite value in phase-4, 'AF' phase, and increases linearly with decreasing of $J_1/|J_2|$ to become equal to 1.2 $k_B/|J_1|$ at $J_1=J_2$.\\

\subsection{\bf Phase $L_4$}
The phase line at $J_1<0$ and $J_2=0$ is located between phase-4 and phase-5 (see Fig. \ref{phasdiag}a). The energy on this line, $u_{L_4}(0)=0$, is confirmed by the simulations as shown in Fig. \ref{area3}a. There is no broken symmetry in this phase and no evidence of a transition from the specific heat simulations.

	This phase has short range order such that no head-to-head orbital pairs occur. This gives a good account of the orbital ordering in a $LaMnO_3$ crystal above its phase transition\cite{Ahmed05}. The entropy in this phase takes the highest value in the whole phase diagram, $s_{L_4} (0) = (0.590\pm0.002)k_B$, which is comparable with that obtained experimentally for $LaMnO_3$ by Sanchez {\it et al}\cite{Sanchez03}. In this case the plot of $T_C$ as a function of $J_2/J_1$ is continuous as shown in Fig. \ref{LMOus}a and takes its minimum value $T_C=0$ on $L_4$.\\
\begin{figure}[h!]
\centering
\includegraphics[scale=0.3]{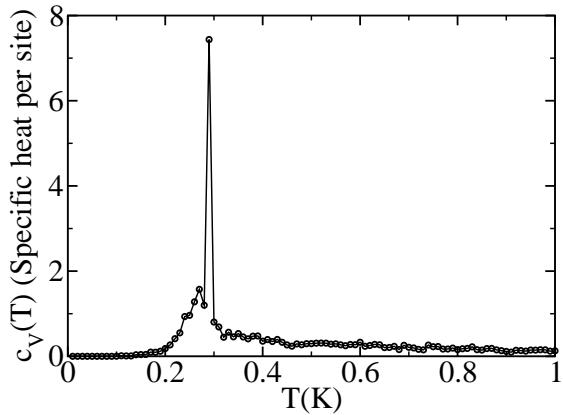}
\caption{\label{L5cv}T-dependence of specific heat, $c_V(T)$, for the line phase $L_5$. It seems to have a strongly first order transition.}
\end{figure}

\subsection{\bf Phase $L_5$}
This is a triple point where three transition lines meet as shown in Fig. \ref{LMOus}a and is found to be strongly first order as seen from the divergence of the specific heat in Fig. \ref{L5cv} compared with that for phase-1. The ground state energy of $L_5$ is the same as for phase-5 and phase-6.
\begin{equation}
u_{L_5}(0)=u_5(0)=u_6(0)=-J_2.
\end{equation}
This phase is a mixed between the two distinct orderings of phase-5 and phase-6. As these phases both have entropies that tend to zero in the thermodynamic limit the entropy is expected to vanish on $L_5$ too. This is confirmed by the simulation with an error equal to $\pm 0.006$. The transition temperature is shown on Fig. \ref{L5cv}. It is seen that it lies on a continuous of the lines from phase-5 and phase-6.

\subsection{\bf Phase $L_6$}
This line separates phase-6 (FM-CB) and phase-1 (FM) phase and occurs for $u_6(0)=u_1(0)$ which is found from Eq. (\ref{uphase1}) and (\ref{u5,6}) to occur for $J_1+J_2=0$. Phase-6 has two transition temperatures. The lower line, $T_{C2}$, goes to zero on $L_6$. The ground state configuration of the phase on $L_6$, as shown in Fig. \ref{lines}d, is ferromagnetic layers alternating with disordered layers. The phase of $L_6$ is called 'FM-Disorder' phase.
\vspace{0.8cm}
\begin{figure}[h!]
\centering
\includegraphics[scale=0.5]{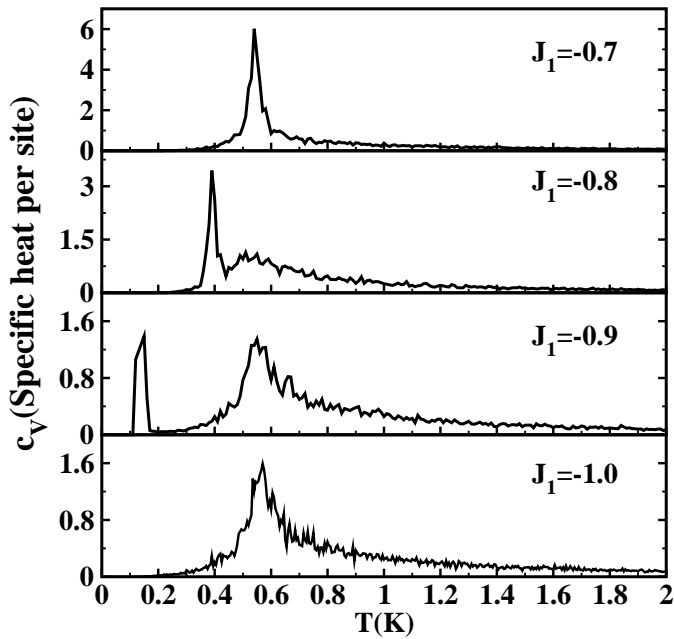}
\caption{\label{cvL6}T-dependence of specific heat, $c_V(T)$, at $J_1=-1, -0.9, -0.8$, and -0.7 and $J_2=1.0$. The upper three panels correspond to phase-1 and the bottom panel is for $L_6$. There exists a short range order peak joining with the transition peak between $L_6$ and $J_1=-0.7$.}
\end{figure}

	The ground state energy can be obtained analytically. The disordered layers have total energy equal to zero but the FM layers are ordered with, say $y$ states in $x$-$z$ plane. So the exchange interaction for each site in a FM layer is $-2J_2$. The total energy at the ground state for the whole lattice is $u_{L_6}(0)=-J_2$. We divided by 2 because only the half of the layers are FM. However, the ground state entropy of the phase on this line comes from the disordered layers which have two states distributed randomly. We predict $s_{L_6}(0)=\frac{k_B}{2} log_e2$, again, we divided by 2 because the entropy comes only from the disordered layers which comprise half of the lattice. The simulated value of the ground state entropy agrees with the analytic one with an error equal to $\pm 0.005$.

	We consider phase-6, phase-1 and the phase on the line joining them, $L_6$, together. The specific heat at the upper transition evolves from a sharp anomaly as shown in Fig. \ref{FM-CB}b to a broader peak at $L_6$ as shown in Fig. \ref{cvL6}. In phase-1 the peak appears to correspond to the onset short range order.

	The fact that we appear to have a line of transition at $T_{C_1}$ in phase-6 that evolves into a short range order peak in the FM phase (phase-1) is reminiscent of the critical end point seen in the liquid-gas transition. However, we see none of the critical phenomena associated with a second order transition. A finite system is unable to distinguish between short range order and long range order below the upper critical temperature on $L_6$. This area of the phase diagram needs further investigations.

\begin{table*}[h!]
\caption{The region and line phases, the analytic ground state energy $u_n(0)$ and entropy $s_n(0)$ per site and transition temperatures $T_C$ obtained with $s_n(0)$ by MC simulation of three states 3d AAFP model for cubic lattice with $L=8$.}
\label{table}
\begin{ruledtabular}
\begin{tabular}{cccccc}
Phase No. &Phase Name     &analy. $u_n(0)$  &analy. $s_n(0)/k_B$  &simulat. $s_n(0)/k_B$  &$k_BT_C$     \\
\hline
phase-1   &FM             &$-J_1-2J_2$  &0               &$0.0\pm0.005$     &---         \\
phase-2   &OFM            &$-J_1-J_2$   &0               &$0.0\pm0.006$     &---         \\
phase-3   &Cage           &$-J_1$/2     &$\frac{1}{4}log_e3$       &$0.270 \pm 0.005$       &---           \\
phase-4   & AF  &0.0 &$\frac{1}{2}log_e2+\frac{1}{32}log_e2$  &0.3673\footnote{Ref (\cite{Wang90})}  &---     \\
phase-5   &LMO             &$-J_2$     &0        &$0.0\pm0.02$    &---          \\
phase-6   &FM-CB           &$-J_2$     &0        &$0.0\pm0.02$    &---            \\
$L_1$ ($J_2=0$)  &RFM&     $-J_1$      &0         &$0.0\pm0.008$     &$0.48J_1$       \\
$L_2$ ($J_2=-J_1/2$)     &Wood Pile   &$-J_1/2$   &$\frac{1}{4}log_e3$   &$0.274\pm0.001$  &$0.39J_1$  \\
$L_3$ ($J_1=0$)    &Disorder        &0        &---     &$0.423\pm0.003$    &---          \\
$L_4$ ($J_2=0$)     &AAFP           &0        &---     &$0.590\pm0.002$    &---          \\
$L_5$ ($J_1=-J_2/2$)    &RFMCB      &$-J_2$     &0     &$0.0\pm0.006$       &$0.29J_2$   \\
$L_6$ ($J_1=-J_2$)    &FM-Disorder     &$-J_2$     &$\frac{1}{2}log_e2$        &$0.342\pm0.005$       &$0.57J_2$  \\
\end{tabular}
\end{ruledtabular}
\end{table*}

	All six regional and line phases obtained pertaining to this phase diagram have been analysed. Two of the phases were the well studied ferromagnetic and antiferromagnetic Potts models with $J_1 =J_2=J$ and $J>0$ and $J<0$ respectively and our results confirmed the known expressions in these cases. One of the phases corresponds to the ordering seen in $LaMnO_3$.

	We used a combination of Monte Carlo simulations and analytic reasoning to obtain our results. The Monte Carlo simulations were run from high to low temperatures and the configurations obtained at the lowest temperatures were analyzed to find if the symmetry had been broken and if so to identify the order parameter. This identified the existence of a phase transition. The ground state energy was obtained from the simulations for all phases. We used the observed configuration to recalculate the ground state energy and in all cases obtained excellent agreement with the simulated values. The analytic expressions for the ground state energies also enabled us to identify the stability lines for each phase by equating the energies of two neighbouring phases.  In all cases the simulations confirmed the phase stability lines that we had found analytically. In most cases knowledge of the ground state configuration enabled us to obtain an analytic expression for the entropy in the ground state. This was harder to obtain from the simulations particularly, the simulations indicated that there was a first order phase transition but again the results of the simulations agreed with the analytic results to within the errors. In the case of two boundary lines there was no ordering and so there was no analytic expression for the entropy and the only estimate was obtained from the simulations.

	The transition temperatures were obtained from the simulations and plots were presented of the variation of the transition temperatures with the variation of the coupling constants.

	It was found that only the ferromagnetic and antiferromagnetic Potts phases were not frustrated in all other cases the ground state energy was higher than the optimal value. A number of novel phases are obtained from our phase diagram. The most unusual results are listed below.

	Phase-3 (or 'Cage' phase) has the most unusual ground state configuration in the phase diagram. Three quarters of the sites are ordered but a large contribution to the ground state stabilization energy comes from the disordered sites. It is seen that the transition temperature falls to zero as $J_1$ is reduced.  The energy stabilization from the disordered sites varies linearly with $J_1$.

	Phase-5 and phase-6 have the same ground state energy but with very different configurations. Their regions of stability were found from considering the free energy at finite temperature which was evaluated using a calculation of the elementary excitations. This was an example of 'order by disorder' that had previously been applied to models with continuous symmetry\cite{Chalker92,Ritchy93}. The method gave the ordering of the phases correctly and also identified the line between them.

	Phase-6 has two well-defined phase transitions. This is very unusual for a Potts model. These phases were found in the simulations and an analytic discussion was given that added understanding. The temperature of the two transitions coincided at the boundary with phase-5 giving a triple point. At the other end of the phase stability the lower transition temperature went to zero at the boundary with the ferromagnetic phase and the upper transition evolved into the onset of short range order.

	In some cases the phase line separated two ordered phases that had various elements in common, an example of that was the transition between ferromagnetism layers stacked antiferromagnetically. In this case the common feature, ordered layers, was preserved on the boundary line and the transition temperature on the boundary line stayed finite and the ground state entropy was zero. In other cases the adjoining phases had no elements in common and in these cases the there was no transition at any finite temperature on the boundary line and a high value for the entropy was found in the low temperature limit.

	In summary this simple Potts model shows a great diversity of phases and different critical behaviour. At least one phase corresponds to a physically realized orbital ordering and it will be interesting to see if any of the other more exotic phases have a physical realization.

\begin{acknowledgments}
This work is funded by the Egyptian High Education Ministry. We would like to thank Prof. K. I. Kugel for reading this paper carefully. M.R.A. thanks Mr. H. Zenia for the fruitful discussions.
\end{acknowledgments}


\end{document}